\documentclass[11pt,a4paper]{article}

\usepackage{tabularx}
\usepackage{graphicx}
\usepackage[small]{caption}

\usepackage{geometry}

\usepackage{amsmath,amsfonts,amssymb}

\newcommand{\be}{\begin{equation}}
\newcommand{\ee}{\end{equation}}
\newcommand{\tr}{{\rm Tr}}

\numberwithin{equation}{section}

\begin{document}

\author{
R.~Lohmayer$^a$, H.~Neuberger$^b$, A.~Schwimmer$^c$, S.~Theisen$^d$\\ [3mm]
{\normalsize\it \llap{$^a$}Institute for Theoretical Physics, University of
  Regensburg}\\
  {\normalsize\it 93040 Regensburg, Germany} \\ [3mm]
  {\normalsize\it \llap{$^b$}Department of Physics and Astronomy, Rutgers University}\\
  {\normalsize\it Piscataway, NJ 08855, U.S.A} \\ [3mm]
  {\normalsize\it \llap{$^c$}Department of Physics of Complex Systems, Weizmann Institute}\\
  {\normalsize\it Rehovot, 76100, Israel} \\ [3mm]
  {\normalsize\it \llap{$^d$}Max-Planck-Institut f{\" u}r Gravitationsphysik, Albert-Einstein-Institut}\\
  {\normalsize\it 14476 Golm, Germany}\\ [7mm]
}

\title{Numerical determination of entanglement entropy for a sphere}

\date{23 November 2009}

\maketitle

\vskip 0.5cm

\begin{abstract}
\noindent We apply Srednicki's regularization to extract the logarithmic term
in the entanglement entropy produced by tracing out a real, massless, scalar field
inside a three dimensional sphere in 3+1 flat spacetime. We
find numerically
that the coefficient of the logarithm is -1/90 to 0.2 percent accuracy,
in agreement with an existing analytical result.
\end{abstract}

\bigskip
\newpage
\tableofcontents
\vskip 1cm

\section{Introduction}

We consider a free, massless, real field $\phi(t,{\vec x})$
defined in four dimensional spacetime,
with $t$ denoting time. We work in the Hamiltonian formalism and
assume that at $t=0$ the system is in its ground state, the vacuum.
We wish to eliminate the quantum
degrees of freedom associated with
$\phi({\vec x})$ and its conjugate momentum $\pi({\vec x})$
located in the spherical region $|{\vec x}| < R$ in space.
We eliminate these degrees of freedom by tracing over all wave functionals
of $\phi({\vec x})$ with $|{\vec x}| < R$. Vacuum
expectation values of operators
${\cal O}$ depending only on $\phi({\vec x})$
and $\pi({\vec x})$ with $|{\vec x}|>R$, denoted as $\phi_{out}$, $\pi_{out}$
respectively, can be expressed with the help
of a density matrix operator $\rho_{out} (\phi_{out}, \phi^\prime_{out})$:
\be
\langle{\cal O}\rangle = \tr {\cal O}\rho_{out}\,.
\ee
$\rho_{out}$ represents a mixed state and a measure of
its ``distance'' from a pure state may be
taken as the von Neumann entropy, S:
\be
S_{out} =-\tr\rho_{out} \log \rho_{out}\,.
\ee
One can trace out the outside degrees of freedom instead,
and obtain $\rho_{in}$, whose entropy
$S_{in}$ is equal to $S_{out}$.

$S_{in}$ is nonzero
because the operators $\phi({\vec x})$ are coupled for
points ${\vec x}$ infinitesimally close to the two sides of the
surface $|{\vec x}|=R$.
Were it not for the spatial
derivative terms in the Hamiltonian, the ground state
would be a single tensor product
over ${\vec x}$ of functionals of $\phi({\vec x})$
and the elimination of the degrees of freedom
inside the sphere would leave a
pure state describing the outside degrees of freedom.
Thus, one can view $S_{in}$ as an entanglement entropy,
where the reference basis is
made out of single tensor products
of functionals of $\phi({\vec x})$. Since the culprit for
$S_{in}\ne 0$ seems localized at the surface of
the sphere, one might guess that $S_{in}$
should depend only on the surface of the
sphere and its embedding in flat spacetime. The
simplest situation would be a flat embedding;
in this paper we deal with the simplest non-flat case.
As the coupling causing the entanglement
occurs at infinitesimal separation,
there is little doubt that a
complete definition of $S_{in}$ will require, at the least,
an ultraviolet cutoff,
a small distance $a$. Without
any cutoff there could be no dependence on $R$ since
the entropy is a pure number. By the same token,
only a logarithmic dependence on $R$ can have an $a$-independent meaning.

The objective of this paper is to extract numerically the coefficient of $\log R$ in
$S_{in}$.

\section{Brief review of previous work}

In~\cite{bomb} a general formula for $S_{in}$ is derived
for Hamiltonians quadratic in
the fields. Only kernels of the type $K({\vec x},{\vec y})$
enter.

After the addition of a mass term to the Hamiltonian, it is shown in~\cite{bomb}
that the entropy per unit surface for a cavity of the form of a three dimensional slab of finite thickness is finite in the $a\to 0$ limit after the subtraction of
a divergent term which does not depend on the thickness.
First ultraviolet and infrared cutoffs are introduced and then
the appropriate limits are taken.

\cite{bomb} also outlines the calculation for more general cavities.
In the spherical case, with massless free fields,
the entropy cannot be finite and $R$-dependent
because $R$ is the single available scale.
The spherical case is somewhat reminiscent of the horizon of a black hole and one may
think of $S$ as a quantum contribution to
the black hole entropy. This makes the sphere particularly interesting.

The spherical case was studied numerically by Srednicki in~\cite{mark}.
Srednicki arrived at the same
setting of the problem as in~\cite{bomb} independently
and took the next step and evaluated
$S_{in}$ for the case of the sphere with a specific regularization.
He found that one only needed to discretize
the spatial radial direction and that there were no infrared divergences.
The short distance structure in the spatial angular directions did not need any ultraviolet
regularization, in conformity with the expectation that it was only the coupling
in the normal direction to the sphere surface that mattered.
If we denote the lattice spacing in the radial direction
by $a$,~\cite{mark} found a leading term in $S_{in}$ as $R/a\to\infty$ which went as $(R/a)^2$.
The coefficient was computed, but it clearly is not a universal number. This was done
for a massless scalar field, so no dimensional parameters beyond $R$ were available
before regularization.

\section{The new result}

We have followed Srednicki~\cite{mark} and pushed his numerical
analysis further, looking for terms in $S_{in}$ that are subleading
in $R/a$. We found subleading terms of the form
 \be
  c \log (R/a) +d\,.
 \ee
We determined the values $c=-1/90$ and $d=-0.03537$ 
with a precision of about two tenths of a percent. 
$d$ is a nonuniversal constant, but the value $-1/90$ 
for $c$ is expected to be a universal number. The coefficient of the 
leading $(R/a)^2$ term also is a nonuniversal number.

Since $c$ might be universal there
ought to be other, analytical, ways to derive it.
An attractive method to do this is based on an
analogue of the ``replica method'', using the identity
\be
S_{in}=-\frac{\partial}{\partial n} \tr\rho_{in}^n \vert_{n=1}\,.
\ee
For integer $n$ one can implement the
trace operation in a Euclidean path integral where
one needs to include a conical singularity reflecting the
spatial sphere. Handling this singularity and, on top of it, the needed
analytic continuation in $n$, makes the application of this method somewhat uncertain. One advantage of this method
is that the universal term can be gotten from the conformal anomaly, perhaps in
closed form and for arbitrarily shaped cavities,
not just a spherical one. Also, one
could envisage an extension to interacting field theory.

In the review~\cite{casini} the result of applying the replica method to the spherical
case is quoted and relevant references are given. The answer they quote~\cite{eqa} is
$c=-1/90$, in agreement with the numerical result of this paper.
In the next section we shall present our numerical work in greater detail, as
the application of the replica trick in conjunction with
conformal anomaly calculations encounters new subtleties
in the case that the surface enclosing the cavity
has extrinsic curvature, as is the case for the sphere~\cite{sch}.

In the 't Hooft large-$N_c$ limit,
for a conformal field theory, one may try to use the
AdS/CFT correspondence in order to calculate the entanglement entropy for various cavities in
the context of strongly interacting conformal field theories.
One needs a prescription for the quantity corresponding to $S_{in}$.
An ansatz that seems to work is reviewed in~\cite{taka}.
This ansatz can be applied to
${\cal N}=4, U(N_c )$ supersymmetric YM theory and produces an entropy
given by $-N_c^2\log R$ for the sphere. If one
uses logarithmic coefficients quoted in
~\cite{casini} for free fields, real scalar (-1/90),
electromagnetic (-62/90) and Weyl fermion (-11/180), one
gets the same value for the logarithmic coefficient at
zero 't Hooft coupling as in the limit of infinite
't Hooft coupling, indicating that this coefficient
is gauge coupling independent in this case. This is consistent
with the view that this coefficient is determined by a non-renormalized
anomaly.

Our numerical work here is a check on one of the numbers that enter the
logarithmic coefficient in the
free case and could be generalized to the other two types of massless fields.
Any general conclusions about the validity of the replica
method, the associated conformal anomaly calculation, and the related AdS/CFT correspondence prescription for entanglement entropy in four dimensions,
in the presence of cavities with surfaces possessing extrinsic curvature,
are left for future work.
More examples might have to be numerically worked out
before matters can be clarified. In this
context, our message is that the accuracy attainable within
reasonable amounts of time on today's consumer-level
desktop computers can suffice in simple enough cases.

\section{Setup of the problem}

Below we summarize the setup of the problem in~\cite{mark}. The Hamiltonian is
\be
H=\frac{1}{2}\int d^3 x [\pi^2({\vec x})+|\nabla\phi({\vec x})|^2 ]\,.
\label{basic}
\ee
$\pi$ and $\phi$ are expanded in spherical harmonics, labeled by integers
$l\ge 0$ and $m=-l,...,l$. This amounts to a canonical transformation to
\be
[\phi_{lm} (x),\pi_{l'm'}(x')]=i\delta_{ll'}\delta_{mm'}\delta(x-x')\,,
\label{commut}
\ee
where $x\equiv|{\vec x}|\ge 0$. The new variables can still be separated into
``inside'' and ``outside'' sets. Now $H=\sum_{lm} H_{lm}$, with
\be
H_{lm} =\frac{1}{2}\int_0^\infty dx\left \{ \pi_{lm}^2 (x) + x^2\left [ \frac{\partial}{\partial x}\left (\frac{\phi_{lm} (x)}{x}\right )\right ]^2 +\frac{l(l+1)}{x^2} \phi_{lm}^2 (x)\right \}\,.
\label{rad}
\ee

The variable $x$ is discretized to $j a$ where $a$ is our short distance cutoff
and $j=1,2,....N$. $N$ is an infrared cutoff which will be taken to infinity.
The range of $l$ is kept infinite and it will be shown that the sum over $l,m$
converges for fixed $N$. This means that one does not need to discretize also
the angular degrees of freedom: no ultraviolet divergences are generated in
the directions tangential to the sphere surface. The finite,
regularized, $H_{lm}$ is:
\be
H_{lm}=\frac{1}{2a}\sum_{j=1}^N\left[\pi_{lm,j}^2+\left(j+\frac{1}{2}\right)^2\left(\frac{\phi_{lm,j}}j-\frac{\phi_{lm,j+1}}{j+1}\right)^2+\frac{l(l+1)}{j^2}\phi_{lm,j}^2\right]\,,
\ee
where $\phi_{lm,N+1}\equiv 0$.
Dropping the $l,m$ indices temporarily, we can write:
\be
H=\frac{1}{2a}\sum_{i,j=1}^N (\delta_{ij}\pi_j^2 +\phi_j K_{ji}\phi_i )\,.
\ee
The real, symmetric, semipositive, tridiagonal $N\times N$ matrix $K$
has non-vanishing entries given by
\begin{align}
K_{11}&=\frac{9}{4}+l(l+1)\,,\\
K_{jj}&=2+\frac1{j^2}\left(\frac{1}{2}+l(l+1)\right)\,, \qquad2\leq j\leq N\,,\\
K_{j,j+1}&=K_{j+1,j}=-\frac{\left(j+\frac12\right)^2}{j(j+1)}\,,\qquad1\leq
j \leq N-1\,.
\end{align}
The radius of the sphere is taken as $R=(n+\frac{1}{2}) a$.

We block decompose $\Omega\equiv\sqrt{K}$:
\be
\Omega=\begin{pmatrix}A&B\\B^T&C\end{pmatrix}\,.
\ee
$A$ is an $n\times n$ matrix with $n<N$. This determines the
dimensions of $B$ and $C$. Let
\be
\beta=\frac{1}{2} B^TA^{-1} B\,,\qquad
\beta^\prime=\frac{1}{\sqrt{C-\beta}}\beta \frac{1}{\sqrt{C-\beta}}\,.
\ee
Then, ~\cite{bomb,mark} the $(N-n)\times(N-n)$ matrix $\beta^\prime$
determines the entropy.
\be
S(n,N)=\sum_{l=0}^{\infty} (2l+1) S_l (n,N)\,,
\ee
where $S_l(n,N)$, the entropy per fixed total angular momentum, is given by
\be
S_l(n,N)=-\tr \left [\log (1-\Xi ) +\frac{\Xi}{1-\Xi} \log \Xi \right ]\,,
\ee
with
\be
\Xi=\frac{\beta^\prime}{1+\sqrt{1-\beta^{\prime 2}}}\,.
\ee
To get $S(n,N)$ we need the eigenvalues of $\Xi$; all square roots and inversions
are well defined and the eigenvalues of $\Xi$, $\xi_j$,
obey $0\le \xi_j \le 1,~j=1,...,N-n$. Srednicki shows that
the sum over $l$ converges at fixed $n,N$ because for $l\gg N > n$ one
has
\be
S_l(n,N)=\xi_l(n)[-\log\xi_l(n) +1]\,,\qquad
\xi_l(n)=\frac{n(n+1)(2n+1)^2}{64 l^2 (l+1)^2}+O(l^{-6})\,.
\label{asyma}
\ee

\section{Numerical details}

The calculation of the $\xi_j$ for any $l$ can be done
in a straightforward manner using {\it Mathematica}.
The choice of {\it Mathematica} is
motivated by its ability to carry out calculations
at arbitrary precision. This facility is costly in computer time for precisions
different from ordinary double float.

One starts by choosing a value of $n$; we find that looking at
values of $n$ in the range of 10 -- 60 suffices for extracting from
$S(n,\infty)$ the term proportional to $\log R$.

We first take the large-$N$ limit at fixed $l$.
Next, the sum over $l$ is performed.
This sum is truncated at a point from where on
the remainder can be done to enough accuracy by employing the large-$l$ approximation~(\ref{asyma}), including also the first subleading term, which we determine numerically (cf. section \ref{sec:inf-l}).

One needs to make sure that the process preserves enough precision.
The ultimate goal is to get $S(n,\infty)$ with an absolute accuracy of $10^{-8}$.

\subsection[The infinite-{$N$} limit]{The infinite-\boldmath{$N$} limit}

By computing $S_l(n,N)$ numerically, we find that for $l\lesssim 15$ the large-$N$ limit of $S_l(n,N)$ is approached like
\be
S_l(n,N)=a_l(n)+\frac{b_l(n)}{N^{2l+2}}\,.
\label{eq:Ndep}
\ee
For higher $l$, it is difficult to determine the exponent of $N$ accurately.
But, it is of the order of $2l$ and therefore finite-$N$ corrections vanish very fast.

\begin{figure}[htb]
\begin{center}
\begin{tabular}{l@{\hspace*{5mm}}r}
    \includegraphics[width=0.46\textwidth]{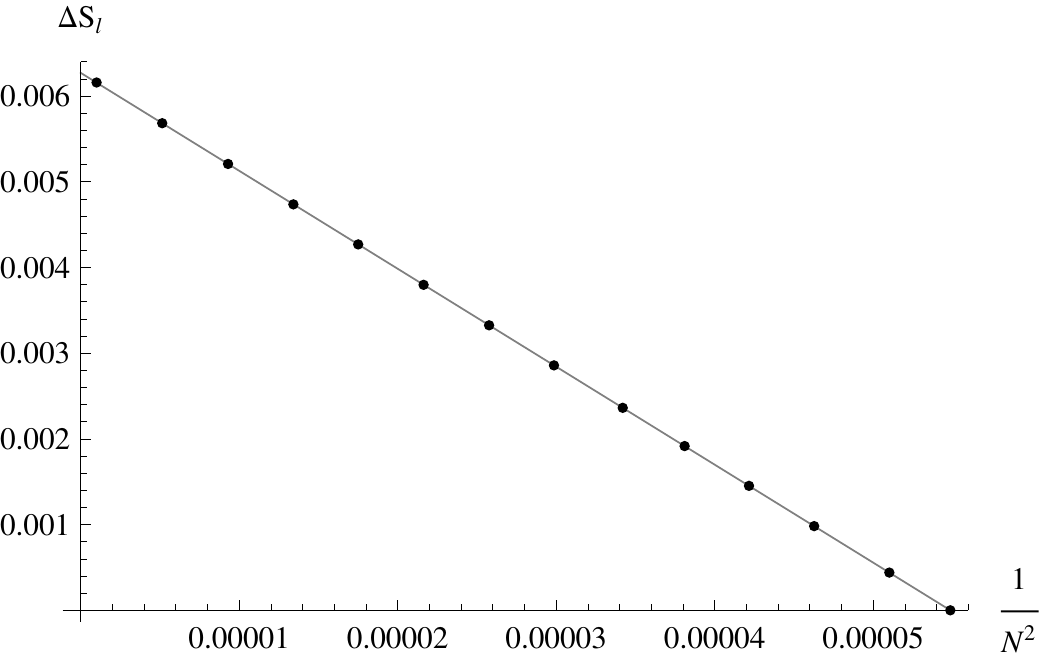} &
    \includegraphics[width=0.46\textwidth]{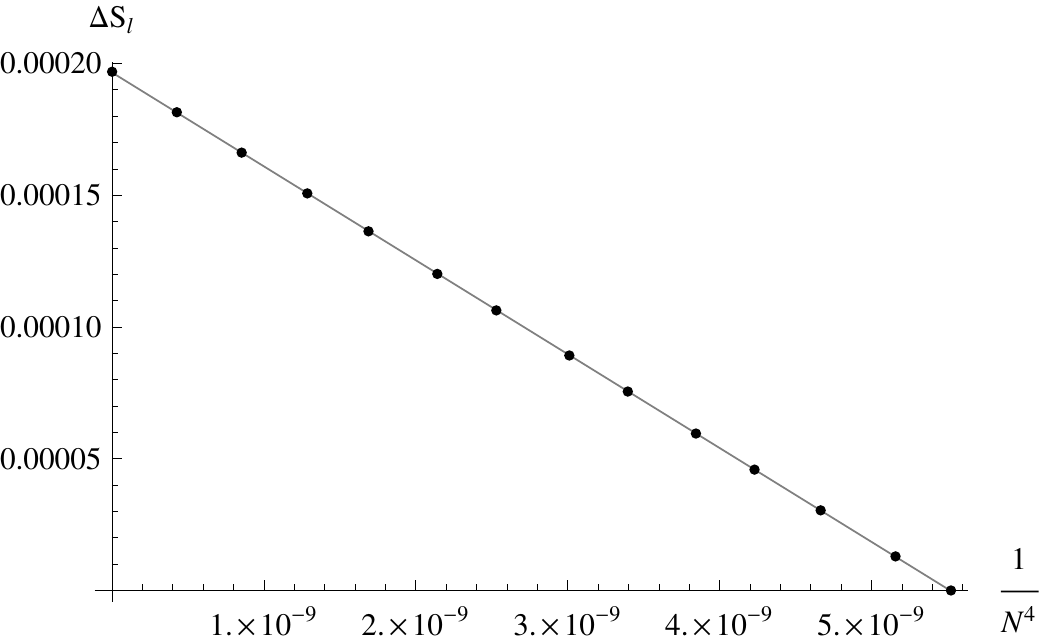} \\
    \includegraphics[width=0.46\textwidth]{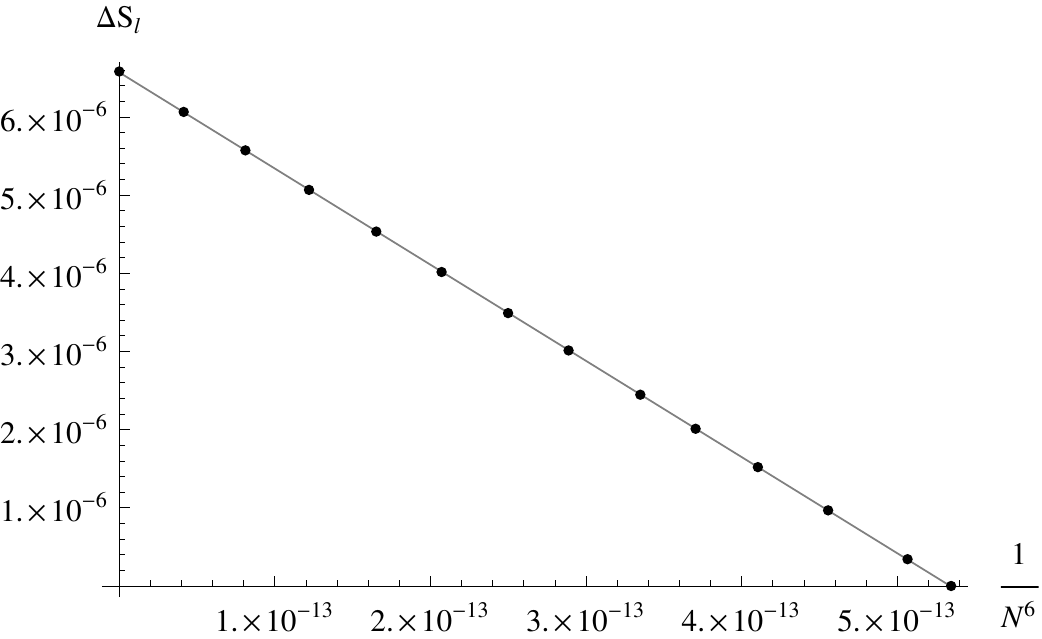} &
    \includegraphics[width=0.46\textwidth]{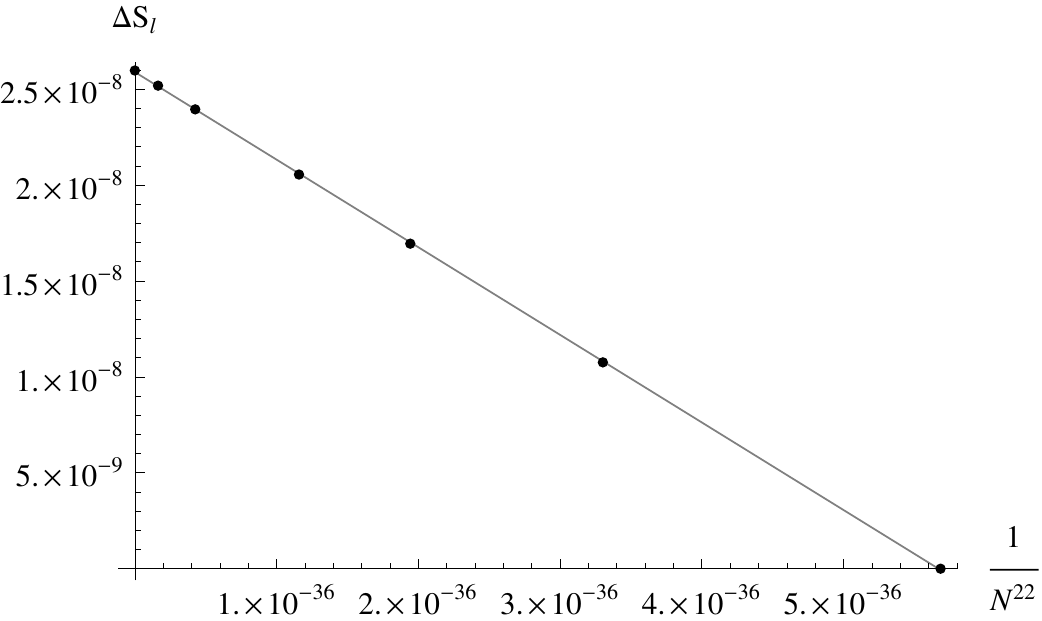}
  \end{tabular}
\caption{Plots of $\Delta S_l(n,N)=S_l(n,N)-S_l(n,N_0)$ as a function of $N^{-2l-2}$ for $n=20$ and $l=0$ (top left), $l=1$ (top right), $l=2$ (bottom left), and $l=10$ (bottom right). The gray lines are straight line fits through the data points (black).}
\label{fig:N-dep}
\end{center}
\end{figure}

$b_l(n)$ was found to be negative in all investigated cases. Figure \ref{fig:N-dep} shows plots of $\Delta S_l(n,N)=S_l(n,N)-S_l(n,N_0)$ as a function of $N^{-2l-2}$ for $n=20$, $l=0,1,2$ and $l=10$ ($N_0$ is the smallest value of $N$ in the data set).

For $n=20$, $l=20$ and $N\geq N_0=60$, $\Delta S_{l=20}(n=20,N)$ is already of the order of $10^{-21}$ (when computed with precision $40$ in {\it Mathematica}).
Only for small $l$ do we have to go to $N$-values as high as a few thousands in order to be able to extrapolate to infinite $N$ with low enough errors.

The cases $l=0,1,2$ and $l\ge 3$ are treated somewhat differently.

For $l=0,1,2$ we extrapolate $S_l(n,N)$ linearly in $1/N^{2l+2}$ to $N=\infty$
applying a least square fit to determine the parameters $a_l(n)$ and $b_l(n)$ in equation (\ref{eq:Ndep}) from evaluations at five high values of $N$.
Varying the number of points used for the fit we obtain estimates for the errors
on the infinite-$N$ limit. See Table 1 for examples. The conclusion is that the
errors are dominated by the $l=0$ contribution. For $l=0$ we have also allowed the
power of $\frac{1}{N}$ to become a fit parameter: this increased the error somewhat
and Table 1 reflects this higher error estimate.

The computation of $S_l(n,N)$ with increased precision in {\it Mathematica} is only possible if $N$ is not too large. The limitation is either the length of time
the computation would take or the available amount of memory.
For small $l$, the extrapolations to infinite $N$ were all performed with $MachinePrecision$. At lower values of $N$, results obtained with $MachinePrecision$ and results computed with increased precision did not differ significantly (between $N=600$ and $N=900$, the relative error is below $10^{-14}$ for $l=0$). Therefore, extrapolations with $MachinePrecision$ are reliable within the estimated error bounds, which do not exceed $10^{-9}$.

For $l\ge 3$
we have carried out full computations
at only two high $N$-values. Based on these numbers we build
various estimates to ensure that even
if the correction for large $N$ goes only as $\frac{1}{N^{2l}}$, rather than
$\frac{1}{N^{2l+2}}$, the large-$N$ limit is still recovered with high
enough precision.

\begin{center}
\begin{table}[htb]
\begin{center}
\begin{tabular}{|l||l|r||l|r||l|r|}
\hline
\multicolumn{1}{|c||}{ $n$ } & \multicolumn{1}{c|}{ $a_0(n)$ } & \multicolumn{1}{c||}{ $\Delta$ }& \multicolumn{1}{c|}{$a_1(n)$} &\multicolumn{1}{c||}{ $\Delta$} & \multicolumn{1}{c|}{$a_2(n)$} & \multicolumn{1}{c|}{$\Delta$} \\
\hline
\hline
10& 0.4779764889& 8e-10 &   0.3218551631505 &1e-13 &   0.24324853242244 & 1.1e-14\\
20& 0.5892154049& 3.0e-9 &  0.4316043027242 &1.9e-12 & 0.35065320435835 & 2.6e-14 \\
30& 0.6553795277& 6.2e-9&   0.4974300032432 &9.9e-12  &0.41592189386854 & 8.8e-14\\
40& 0.7026231983& 9.9e-9 &  0.544542664951 &3.0e-11 &  0.46281476449290 & 3.4e-13\\
50& 0.739392926& 1.3e-8 &   0.581247646267 &6.9e-11 &  0.4994098682251 &  1.1e-12 \\
60& 0.769499767& 1.6e-8 &   0.61131758758 & 1.3e-10 &  0.5294166892935 & 3.0e-12 \\
 \hline
\end{tabular}
\end{center}
\caption{Results of extrapolations to infinite $N$ (and corresponding absolute error bounds $\Delta(n,l)$) for $l=0,1,2$. Ranges of $N$ used to extrapolate: $3000\leq N \leq 5000$ for $l=0$, $1500\leq N \leq 3000$ for $l=1$ and $1000\leq N \leq 2000$ for $l=2$. For a fixed set of values of $N$ at which full calculations are made,
the error decreases with increasing $l$. Although smaller $N$ values are used
for $l=1,2$, the estimates on the errors in these cases
are smaller than those for $l=0$.}
\label{tab:lowl}
\end{table}
\end{center}

\subsection[The infinite sum over $l$]{The infinite sum over \boldmath{$l$}}
\label{sec:inf-l}
Having taken the infinite-$N$ limit for all finite $l$'s,
we now turn to performing the infinite sum over $l$. For every $n$, we can
compute, as described above, the value of $S_l(n,N=\infty)$. We do this
for $l=0,1,2,...l_{\text{max}}$ and then use the leading term in
equation~(\ref{asyma}) to estimate the remainder of the sum, stemming from
contributions starting at
$l=l_{\text{max}}+1$ and all the way to $l=\infty$.
This procedure can be further improved on, by doing some
calculations at a few selected very high values of $l$
and looking at the difference between the
leading asymptotic form and the numerical result. In this way we get
an assessment for the subleading term in~(\ref{asyma}).
By this method we convince
ourselves that the values of $l_{\text{max}}$ we end up
using in conjunction with the asymptotic result
provide an absolute accuracy on the final numbers of order $10^{-8}$.

\section{\boldmath Asymptotics at large $R$}

\begin{figure}[htb]
\begin{center}
    \includegraphics[width=0.75\textwidth]{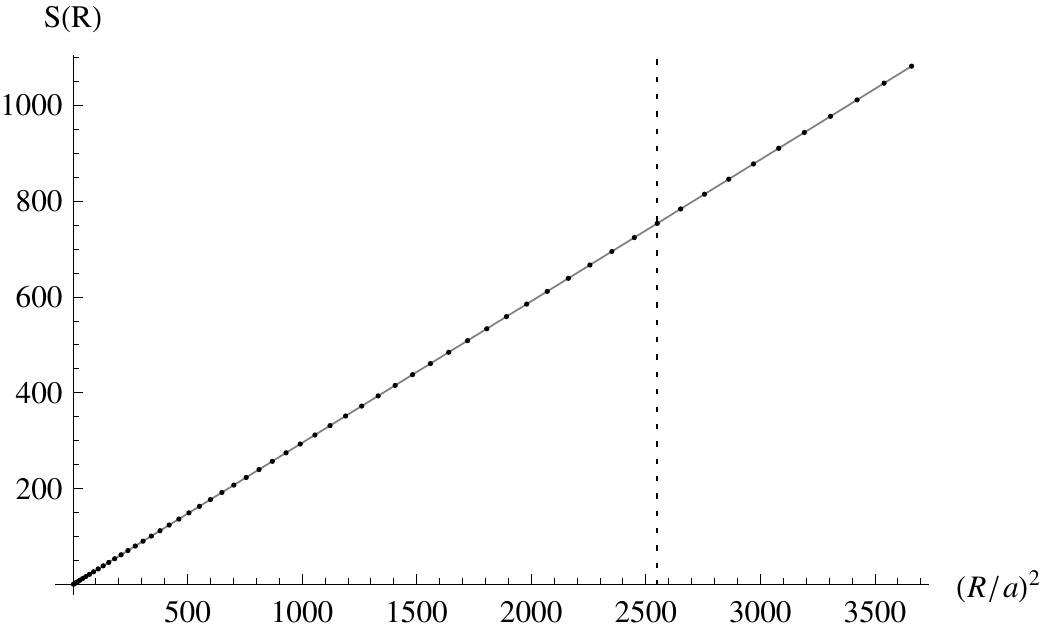}
  \caption{Plot of $S(R)$ as a function of $(R/a)^2$, the gray line is a fit through the last $20$ points.}
\label{fig:LinRes}
\end{center}
\end{figure}

We end up with a set of numbers for $S(R)$ for $R^2$ up to about $3700a^2$.
These numbers vary from order one to a few hundreds and are accurate
to about $10^{-8}$, that is to at least eight digits.
\be
S(R=(n+1/2)a)\equiv\lim_{N\to\infty}S\left(n,N\right)=\lim_{N\to\infty}\sum_{l=0}^\infty (2l+1) S_l(n,N)\,.
\ee
Figure \ref{fig:LinRes} shows a plot of $S(R)$ as a function of $(R/a)^2$, confirming the area law found in \cite{mark}. The gray line through the data points (obtained from a fit through the last 10 points, to the right of the vertical dashed line) is given by
\be
S_{\rm{lin}}(R)=0.295406\, (R/a)^2\,.
\ee
Srednicki quotes a slope of 0.30, so we confirm the two digits he has found.

Next, we fit the data points to the functional form
\be
S_{\rm{log}}(R)=s (R/a)^2+ c' \log (R/a)^2+d\,.
\label{eq:Slog}
\ee
A least square fit over the last 15 data points, $45\leq n\leq 60$ results in
\be
s=0.295431\,,\qquad c'\equiv c/2 = - 0.005545\,,\qquad d=-0.03537\,.
\ee
Note the change in $s$, by $2.5\cdot 10^{-5}$.
When we change the range of data points used in the fit, the result for $s$ does not change to the given precision, variations in $c'$ are of the order $10^{-5}$ and variations in $d$ are of the order $10^{-4}$. The higher error in $d$ indicates that
the least square fit altered the coefficient
of the log somewhat, away from its
true value (which could have been
obtained if we would have fit even further
subleading terms). But, three to four digits accuracy is very likely.

\begin{figure}[htb]
\begin{center}
    \includegraphics[width=0.75\textwidth]{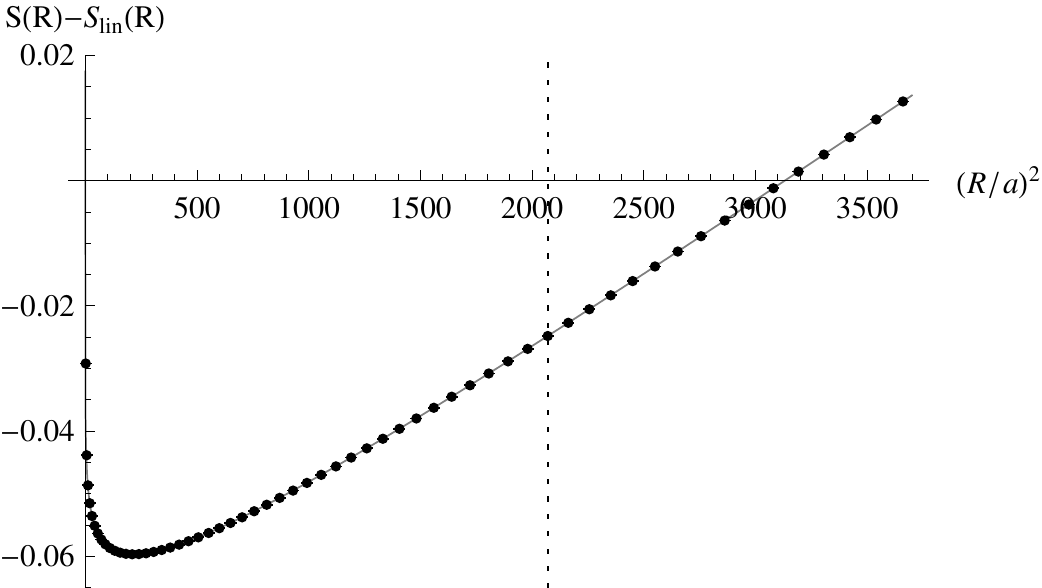}
   \caption{Plot of $S_{\rm{log}}(R)-S_{\rm{lin}}(R)=2.5\cdot10^{-5}(R/a)^2-0.005545 \log (R/a)^2-0.03537$ as a function of $(R/a)^2$ (solid gray curve) and computed data points $S(R)-S_{\rm{lin}}(R)=S(R)-0.295406 (R/a)^2$  (black dots). $S_{\rm{log}}(R)$ is obtained from a fit over the last 15 data points, to the right of the vertical dashed line. Error bounds are not visible, being of the order $10^{-8}$.}
   \label{fig:DiffLinRes}
\end{center}
 \end{figure}

Figure \ref{fig:DiffLinRes} shows a plot of the difference between the two fits, $S_{\rm{log}}(R)-S_{\rm{lin}}(R)$, as a function of $(R/a)^2$, and the corresponding data points.

\section{Universality}

Equations~(\ref{commut}) and~(\ref{rad}) are somewhat formal because of
the need to regularize. At this level one can make a formal canonical
transformation and afterward discretize the resulting expression.
{\it A priori} there is no guarantee that the coefficient $c$ of
$\log R$ in the entropy will again come out as -1/90.

Let us sketch an example which works in the same way in every $(lm)$ sector;
then, for brevity, we can drop the $l,m$ indices. We make the canonical
transformation
\be
q(\xi)=\phi(e^\xi ),~~p(\xi)=e^\xi \pi(e^{\xi})\,.
\label{cantrans}
\ee
We have
\be
[q(\xi),p(\xi')] = i\delta(\xi-\xi')\,.
\ee
In terms of $q,p$, we get:
\be
H=\frac{1}{2}\int_{-\infty}^{\infty} d\xi e^{-\xi} \left [ p^2(\xi)
+ \left (\frac{dq}{d\xi} - q(\xi) \right )^2 + l(l+1) q^2(\xi) \right ]\,.
\ee
A further canonical transformation, with $p(\xi)\to p(\xi-\xi_0)$,
$q(\xi)\to q(\xi-\xi_0)$ results in $H\to e^{\xi_0} H$.
The entropy depends only on the ground state of $H$, so remains
invariant under a rescaling of $H$ by a positive number. It now seems that we
can always absorb $R$ in $\xi$ and $S$ will be $R$-independent.

If we want to preserve the simple behavior of $H$ under $\xi\to\xi-\xi_0$
we would need to discretize $\xi$ on a regular lattice,
$\xi\to j a,~-\infty < j<\infty$.\footnote{The term $e^{-\xi}$ in the density
raises concerns about the region close to the center of the sphere,
but we could taper off the dependence of $\xi$ on $j$
for values of $j\ll 0$ to something more manageable.
Note that this has to do with the vicinity of the center
of the sphere, far from its surface.
This would break the $\xi$-shift symmetry somewhat. }
It is not easy to see how exactly the various limits will work out,
but one might conclude that the entropy does not depend
on $R$ at all.

The set of regularizations under which the coefficient of
$\log R$ is fixed at -1/90 must then be, at the least, restricted by some
additional requirements.
Assuming that $c$ is indeed determined by an anomaly, it becomes apparent,
as is always the case with anomalies, that the true consequence of
being forced to employ a regularization
is that there are several symmetries which
cannot be simultaneously preserved in the quantum continuum limit.
If we insist on maintaining scale invariance, some other symmetry will
have to be violated. The most likely candidate in our example is
the ${\vec x}\to {\vec x}-{\vec x}_0$ three dimensional
translational invariance of~(\ref{basic}). Although broken at finite
spacing $a$ in Srednicki's regularization, we would guess that in this case it gets restored as $a$ goes to $0$.

To us it seems likely that requiring three dimensional translational invariance
in the continuum limit would fix $c$ to -1/90. Allowing this invariance to break may
produce different values for $c$, among them even 0 if scaling becomes
fully preserved in the continuum limit.

To be sure, we have certainly not shown this
here. Substantially more work would be needed to produce a convincing
numerical argument for the universality of $c$ and its limitations.

The replica method turns the evaluation of $S$ into a calculation of
the partition function of a four dimensional Euclidean field theory
consisting of a free scalar field on a manifold which has a conical
singularity. The dependence on $R$ can then be extracted from the
variation of the free energy with respect to a background four
dimensional metric. Classically the free energy is invariant under
Weyl transformations and diffeomorphisms. Quantum mechanically both
symmetries cannot be maintained  simultaneously and  an anomaly (the
"Weyl anomaly") could appear. To connect to the procedure we use in
this paper we should study how the above mentioned symmetries are
implemented through canonical transformations.

For the Weyl transformations we define  canonical transformations of
the fields defined in \eqref{commut} parametrized by a function
$\sigma(x)$, depending just on the radial coordinate:
\begin{align}
\phi(x)&\to\tilde\phi(x)=\phi(x)e^{-\sigma(x)}\,,\\
\pi(x)&\to\tilde\pi(x)=\pi(x)e^{\sigma(x)}\,,
\end{align}
where we have again dropped the $l,m$ indices. It is clear that the
above canonical transformation can be discretized by setting $x=ja$.
It is also obvious that $S(n,N)$ will not change if we first
carry out the canonical transformation. Therefore, $c$ would not depend
on the set of numbers $\sigma_j,~j=1,...,N$.

For the diffeomorphisms, on the other hand, we define the canonical
transformations
 \be
q(\xi)=\phi(\tau(\xi)),~~p(\xi)=\tau'(\xi)\pi(\tau(\xi))\,,\label{canb}
\ee
parametrized by an arbitrary monotonic function $x=\tau(\xi)$,
generalizing (\ref{cantrans}). Under (\ref{canb}) the Hamiltonian is
replaced by
 \be H=\frac{1}{2}\int d\xi\frac{1}{\tau'(\xi)} \left[
p^2(\xi) + \left(q'(\xi)-\frac{\tau'(\xi)}{\tau(\xi)}q(\xi)\right)^2
+ l(l+1)\left(\frac{\tau'(\xi)}{\tau(\xi)}\right)^2  q^2(\xi)
\right]\,. \ee
Discretizing  $\xi$, we could repeat the numerical
procedure and study the  dependence of the coefficient $c$, if any,
on the discrete values $\tau_1,\tau_2,...$ parametrizing the
function $\tau(\xi)$. In principle, one can decide if the
dependence survives in the continuum limit.

In the four dimensional path integral obtained via the replica
method, a Weyl anomaly appears. Specifically, \cite{sch} suggests
that the anomaly is intrinsically four-dimensional, implying for our
case that $c$ depends on the entire function $\tau(\xi)$. On the
other hand, the ansatz of \cite{taka} predicts that the anomaly is
two-dimensional, i.e. $c$ does not depend on $\tau(\xi)$.

\section{Conclusion}

Our study leads us to the conclusion that the logarithmic term in the entanglement
entropy associated with a spherical cavity of radius $R$, in the case of a free
massless scalar field in four dimensions is
$\Delta S (R) = c\log R/a$
with $c=-1/90$ within an error of 0.00002.

We have already commented in previous sections how
this number compares to analytical
results and how general its determination is.
The experience gathered while carrying out this
exercise should be useful for future numerical work.

\subsection*{Acknowledgments.}

RL and HN acknowledge partial support by the DOE under grant
number DE-FG02-01ER41165. RL acknowledges support by BayEFG.
AS and ST acknowledge The German-Israeli Project Cooperation (DIP H52),
the Einstein Center of the Weizmann Institute and the Humboldt Foundation
for partial support.
HN notes with regret that his research has for a long time been
deliberately obstructed by his high energy colleagues at Rutgers.
AS and ST acknowledge David Kutasov for discussions.

\end{document}